The following document is the final published version of an article that was accepted and published in the open access journal "*Bioelectromagnetics*":

# The Importance of Subcellular Structures to the Modeling of Biological Cells in the Context of Computational Bioelectromagnetics Simulations

Kevin Jerbic ,[1]* Jan T. Svejda ,[1] Benedikt Sievert ,[1] Andreas Rennings ,[1] Jürg Fröhlich ,[2] and Daniel Erni [1]

[1]*General and Theoretical Electrical Engineering (ATE), Faculty of Engineering, University of Duisburg-Essen, and Center for Nanointegration Duisburg-Essen (CENIDE), Duisburg, Germany*
[2]*Fields at Work GmbH, Zürich, Switzerland*

Numerical investigation of the interaction of electromagnetic fields with eukaryotic cells requires specifically adapted computer models. Virtual microdosimetry, used to investigate exposure, requires volumetric cell models, which are numerically challenging. For this reason, a method is presented here to determine the current and volumetric loss densities occurring in single cells and their distinct compartments in a spatially accurate manner as a first step toward multicellular models within the microstructure of tissue layers. To achieve this, 3D models of the electromagnetic exposure of generic eukaryotic cells of different shape (i.e. spherical and ellipsoidal) and internal complexity (i.e. different organelles) are performed in a virtual, finite element method-based capacitor experiment in the frequency range from 10 Hz to 100 GHz. In this context, the spectral response of the current and loss distribution within the cell compartments is investigated and any effects that occur are attributed either to the dispersive material properties of these compartments or to the geometric characteristics of the cell model investigated in each case. In these investigations, the cell is represented as an anisotropic body with an internal distributed membrane system of low conductivity that mimics the endoplasmic reticulum in a simplified manner. This will be used to determine which details of the cell interior need to be modeled, how the electric field and the current density will be distributed in this region, and where the electromagnetic energy is absorbed in the microstructure regarding electromagnetic microdosimetry. Results show that for 5 G frequencies, membranes make a significant contribution to the absorption losses. Bioelectromagnetics. 44:26–46, 2023. © 2023 The Authors. *Bioelectromagnetics* published by Wiley Periodicals LLC on behalf of Bioelectromagnetics Society.

## INTRODUCTION

As part of the network expansion towards 5 G, depending upon future technological or regulatory changes, people may become increasingly exposed to electromagnetic (EM) radiation at higher frequencies (24–52 GHz) compared to those currently implemented by telecommunication standards (2 G at 1.9 GHz, 3 G at 2.1 GHz and 4 G from 600 MHz up to 2.5 GHz). This trend is accelerated by ambitious infrastructure projects such as Industry 4.0 or the Internet of Things. This means that electromagnetic exposure will not be limited primarily to public spaces, as has been the case in the past, but will also have a massive impact on the workplace or remote

Grant sponsor: Swiss Research Foundation for Electricity and Mobile Communications (FSM), grant number: A2019-01; Deutsche Forschungsgemeinschaft (DFG, German Research Foundation), grant number: Project 287022738; Konzepte für Ende-zu-Ende Terahertz 6G-Mobilfunk, grant number: no. 16KISKK039.

*Correspondence to: Kevin Jerbic, General and Theoretical Electrical Engineering (ATE), Faculty of Engineering, University of Duisburg-Essen, and CENIDE – Center for Nanointegration Duisburg-Essen, D-47048 Duisburg, Germany.
E-mail: kevin.jerbic@uni-due.de

Received for review 24 June 2022; Revised 15 November 2022; Accepted 28 January 2023

DOI:10.1002/bem.22436
Published online 16 February 2023 in Wiley Online Library (wileyonlinelibrary.com).

areas such as those used for agriculture implementing smart farming. The resulting increase in the duration of exposure to EM fields at frequencies that are currently absent or only present at very low extents in our environment necessitates a more detailed reassessment of existing radiation guidelines, as anticipated by the International Commission on Non-Ionizing Radiation Protection (ICNIRP) [2020]. This shift to higher frequencies thereby brings into focus the absorption in the outermost millimeters of the body surface, especially in the skin and the eyes, and raises the question of exactly where in a tissue microstructure the energy absorption occurs. Investigating the effects of the interaction between EM fields and biological systems down to the (sub)cellular level, therefore, requires high-resolution microscopic tissue models.

In order to perform microdosimetric studies to predict these effects of interaction and, in particular, the EM energy intake into tissues and their underlying cells, accurate cell models are needed. However, the development of such models is challenging because cells are highly complex symbiotic systems consisting of a multitude of organelles stochastically distributed inside them [Kitano, 2002]. For this reason, the level of detail used in a model is defined by the most relevant structures significantly influencing the field distribution regarding the EM field quantities of interest. This implies that the suitability of a cell model depends on the purpose for which it is used.

In its simplest representation, a eukaryotic cell is assumed to be a concentric shelled sphere or ellipsoid. In this representation, the interior of the cell is divided into two distinct compartments: the cytoplasm (CP) and the nucleus. These are separated from each other and from the exterior by thin membranes. Although this assumption (over)simplifies reality, analytical studies using such cell models have been successfully employed to investigate the distribution of EM fields in the cell interior, allowing the approximate determination of induced membrane voltages and losses [Kotnik and Miklavčič, 2000, 2006; Vajrala et al., 2008; Ye et al., 2010]. This simple representation of a cell has also proven useful in approximating the EM material properties of the two compartments and the membranes under consideration. This was performed in two steps. First, dielectric spectroscopy was used to determine the frequency-dependent macroscopic dielectric properties of cell suspensions [Asami et al., 1979, 1989; Lisin et al., 1996; Polevaya et al., 1999]. In a second step, by inversely applying mixing rules to these properties assuming the same model as used above, a representation of the cell and its compartments was created that corresponded to the statistical average of all cells contained in the suspension allowing their material properties to be derived [Ermolina et al., 2000; Ermolina et al., 2001; Asami, 2002; Feldman et al., 2003; Sihvola, 2008; Merla et al., 2009].

Recent developments in cell modeling have been influenced by improvements in computer hardware and software. Using numerical methods, such as the finite element method (FEM), it is possible to calculate the field distribution within cell models that have a much higher morphological complexity, which is expressed in both their shape and internal organization. This allows for a variety of numerical studies that consider the interaction of EM fields with eukaryotic cells in different contexts and fields of application. Focusing on the effects of cell shape on both the EM field distribution and effective macroscopic material properties, such studies have conducted numerical investigations using Giehl's superformula [Gielis, 2003] simulating a variety of human cells in vitro [Mescia et al., 2018] and in vivo [Huclova et al., 2010; Dermol-Cerne et al., 2018]. Another trend in cell modeling has been initiated by the advent of biosensors that allow dielectric spectroscopy on single cells. In combination with the numerical methods mentioned above (i.e. FEM), not only is it possible to consider the interaction of EM fields with cells in a generic setting but also to create digital twins of underlying experiments, thus allowing phenomenological interpretation of the measured spectral responses [Merla et al., 2018; García-Sánchez et al., 2018; Feng et al., 2019; Merla et al., 2019; Guo et al., 2021]. Due to the high degree of correlation required between such simulations and their corresponding experiments, fluorescence microscopy is used alongside edge detection and extraction algorithms to incorporate the individual shape of the plasma membrane (PM) as well as that of the internal organelles of the cells under investigation. Two and three-dimensional models generated in this manner are introduced in Denzi et al. [2016, 2017, 2018], de Angelis et al. [2019] and [Pucihar et al., 2006; Towhidi et al., 2008; de Angelis et al., 2020], respectively.

Given this state of the art, cell models used to address risk assessment [IEEE, 2019; ICNIRP, 2020] require a volumetric representation of eukaryotic cells and their organelles. In order to determine which are the relevant details that have to be modeled, the present study systematically investigates and compares different cell models of increasing levels of detail. The cell models under consideration start from the simple spherical cell model mentioned above, containing only a nucleus, and increase in complexity to an ellipsoidal cell comprising a fully resolved endoplasmic reticulum (ER) surrounded by a membrane of low conductivity. Special attention has been paid to the volumetric representation of organelles within the cell. The ER in the most complex cell model, for instance, is conceptualized as a distributed membrane system, and its volumetric extent is approximated by cisternae concentrically surrounding the nucleus. Two methods are therefore presented here to determine the effect of gradually increasing complexity on the resulting field distribution. One of these methods allows the qualitative separation of and distinction between the influence of dispersive material properties and changes in the geometric structure of the cell. The other allows the quantification and graphical illustration of the effect of an ER on the heterogeneity of the EM field distribution in the interior of the cell. Considering all of these details, the following questions will be answered: (1) How is the current and volumetric loss density distributed in the cell interior? (2) Where is EM energy absorbed? (3) What are the effects of different levels of detail of the underlying cell models on the resulting macroscopic material properties? At all levels of complexity, the EM field distribution within these models is numerically determined using the FEM in the context of a virtual capacitor experiment in the frequency range between 10 Hz and 100 GHz.

To address all of the above, the methodology is presented in the second section, including a detailed description of the eukaryotic cell models of different complexity, their dispersive material properties, and the computational setup in which they are placed. In addition, special focus is given to the introduction of the two methods outlined in the previous paragraph. Following this, the cell models are investigated with respect to their internal field distributions and effective macroscopic material properties. The results of these investigations are presented in the third section. Finally, these results and their consequences regarding future research are discussed in the conclusion (fourth section) and an outlook (fifth section). As an addition, the main study is followed by an appendix where the procedure for applying mixing rules in order to verify the results is provided in closer detail.

## METHODOLOGY

The methodology is divided into four subsections. In the first of these, the eukaryotic cell models under investigation are introduced alongside a method used to spatially resolve and decompose the cell interior for microdosimetric investigations focusing on current and loss distributions within the cell models. Following on from that, the dispersive material properties of the individual cell compartments are derived based on published data. In the third subsection, a simulation setup is presented to perform a FEM-based capacitor experiment to predict the distribution of EM fields within the cell, and finally, in the last subsection, the investigations conducted are introduced.

### Eukaryotic Cell Models and Spatial Segmentation

The three generic cell models investigated in this study are depicted in Fig. 1a–c: (i) a simple spherical cell containing a spherical nucleus, (ii) an ellipsoidal cell containing a spherical nucleus, and (iii) an ellipsoidal cell containing both a spherical nucleus and a distributed ER. The so-called rough ER is approximated by cisternae which surround the nucleus concentrically.

Taking these structures into account, the cell interior can be divided into two or three compartments: the CP, the nucleoplasm (NP), and the ER, each separated from the other and from the exterior surrounding, i.e. extracellular medium (EC), by a lipid membrane of low conductivity. The subdivision of the cell interior into separate (computational) (sub) domains allows field distributions to be determined with a spatial resolution that allows current and volumetric loss densities to be integrated in each of these (sub)domains. The volumetric segmentation is shown as an illustrative example for the most complex cell model in Fig. 1d–f. In addition, the (cross-) sectional segmentation of the cells is shown in Fig. 1g–i allowing the determination of EM fields in the XY, YZ, and XZ planes. Using these sectional planes in combination with the volumetric segmentation of the cell into its compartments, it is possible to determine the currents flowing through each of them, as shown in Fig. 1j–l.

The simple spherical model shown in Fig. 1a is based on lymphocytes. Lymphocytes were chosen as a template for an initial cell model for two reasons: (i) Their dielectric properties have been extensively studied by the dielectric spectroscopy community [Ermolina et al., 2001; Feldman et al., 2003]. (ii) They have a very uniform round shape and their internal structure is very well documented, allowing the study of membrane properties on a cell geometry of low complexity. According to Asami et al. [1989], their nucleus occupies about 40 %–70 % of the cell volume and the CP surrounding the nucleus is poor in organelles. Due to these physiological properties, this first model is well suited for the verification of bottom-up multiscale modeling using mixing rules [Ermolina et al., 2000]. In this study, however, the spherical cell model is investigated with respect to the size of the nucleus using numerical simulations. The ratio of the volume of the nucleus to that of the whole cell $c_N$ normally ranges from 0.4 to 0.7, but here has been extended to cover 0.3–0.9 in order to investigate more extreme cases. Following [Ermolina et al., 2000], the cell radius was chosen to be $r_{cell} = 7$ μm, resulting in minimum and maximum radii of $r_{N,min} = 4.69$ μm and $r_{N,max} = 6.76$ μm derived from the range given above using $r_N = \sqrt[3]{c_N} \cdot r_{cell}$. In addition, the reference nuclear radius is $r_{N,ref} = 5.9$ μm corresponding to $c_N = 0.6$.

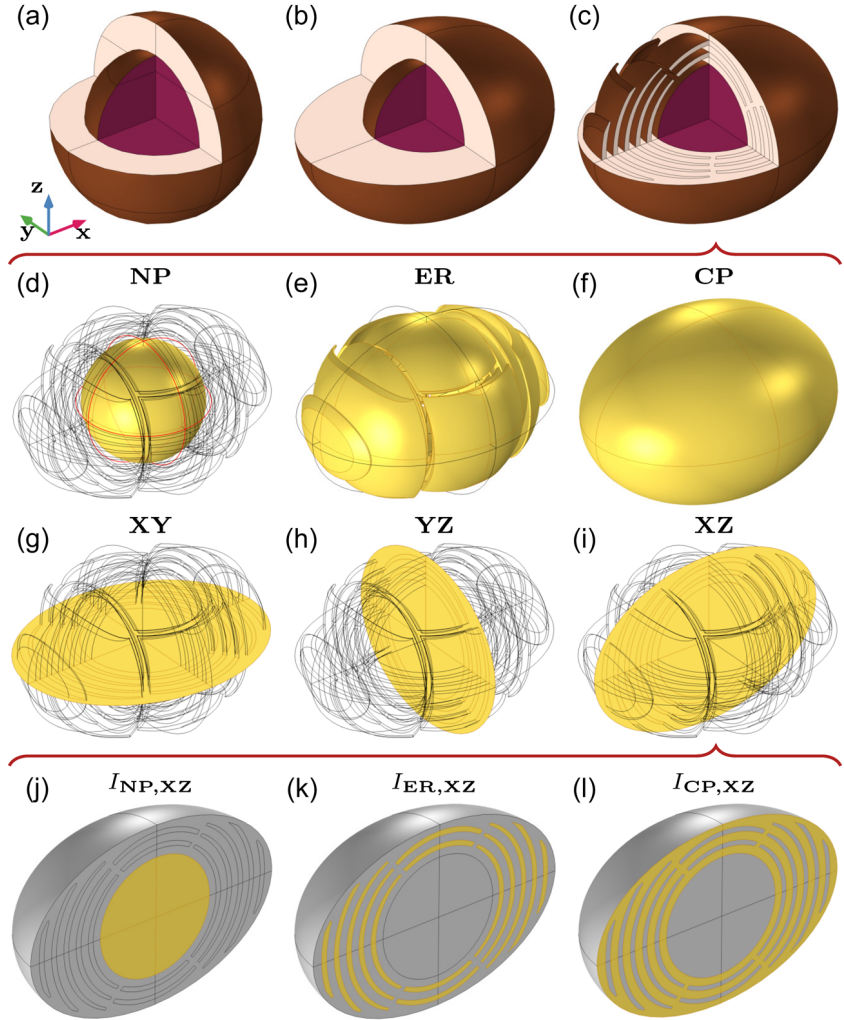

Fig. 1. Overview of the cell models studied and the segmentation used for postprocessing: (**a–c**) Generic eukaryotic cells of different shapes and internal complexity. Starting with a simple spherical cell model containing only a nucleus (N) in (**a**), the level of complexity increases to an ellipsoidal cell model comprising a fully resolved endoplasmic reticulum (ER) in (**c**). (**d–f**) Segmentation of the cell interior into (computational) domains to capture integral quantities such as the corresponding compartment losses. In the given example, the domains highlighted correspond to the nucleoplasm (NP), the ER and the cytoplasm (CP). (**g–i**) Sectional planes for the evaluation and integration of internal fields showing the XY, YZ, and XZ planes, respectively. (**j–l**) Segmentation of the sectional planes to capture integral quantities such as the corresponding compartment currents. In the given example, the areas correspond to the currents in the NP, the endoplasmic reticulum and the CP in the XZ plane. The areas of the sectional planes corresponding to the integration are highlighted.

In order to investigate the influence of compartmentalization within the cell, the simple spherical cell model is now extended to a more realistic approximation of a cell in the following two steps. Keeping the nuclear radius at a value of $r_{N,ref} = 5.9$ μm, which was previously considered as the reference radius, the PM is first transformed into an ellipsoidal one with semi-axes $r_{cell,x} = 13\,\mu m$, $r_{cell,y} = 10.5\,\mu m$, and $r_{cell,z} = 9\,\mu m$ (cf. Fig. 1b). The influence of this change of shape on the macroscopic material properties and the microdosimetric measures is investigated with respect to the resulting anisotropy. To capture the current density and volumetric loss density distribution within the cell in the context of more complex compartmentalization, a distinct ER was added to the cell in a second step, allowing the cell to be studied as a distributed membrane system (cf. Fig. 1c). The ratio of the volume of the ER to the total cell volume is $c_{ER} = 0.2$, that of the CP is $c_{CP} = 0.63$ and that of the nucleus, $c_N = 0.17$. The thickness of the cell membrane and that of the elements of the ER is $t_M = 7\,nm$ and that of the nuclear envelope is $t_{NE} = 40\,nm$. To provide a better overview, the geometrical properties are summarized in Table 1.

## Derivation of the Dispersive Material Properties Based on Published Data

All cell components have dispersive material properties which are modeled using a lossy Debye-type frequency dependence:

$$\underline{\varepsilon}_r(\omega) = \varepsilon_{r,\infty} + \frac{\varepsilon_{r,s} - \varepsilon_{r,\infty}}{1 + j\frac{\omega}{\omega_0}} - j\frac{\sigma_s}{\varepsilon_0 \omega} \qquad (1)$$

were $\varepsilon_{r,\infty}$ and $\varepsilon_{r,s}$ denote the permittivity at the high frequency and the low-frequency limit of the underlying polarization mechanism respectively. $\sigma_s$ stands for the DC conductivity of the material.

**TABLE 1. Geometric Specifications of the Cell Models Under Investigation**

|  | SC | EC | ER | Description |
|---|---|---|---|---|
| $r_{cell}$ | $7\,\mu m$ | – | – | Radius of the sph. cell model |
| $r_{cell,x}$ | – | $13\,\mu m$ | $13\,\mu m$ | Semi-axis in x-direction |
| $r_{cell,y}$ | – | $10.5\,\mu m$ | $10.5\,\mu m$ | Semi-axis in y-direction |
| $r_{cell,z}$ | – | $9\,\mu m$ | $9\,\mu m$ | Semi-axis in z-direction |
| $r_{N,(ref)}$ | $(5.9\,\mu m)$ | $5.9\,\mu m$ | $5.9\,\mu m$ | Radius of the nucleus |
| $t_M$ | $7\,nm$ | $7\,nm$ | $7\,nm$ | Thickness of the membranes |
| $t_{NE}$ | $40\,nm$ | $40\,nm$ | $40\,nm$ | Thickness of the nuc. envelope |
| $c_{N,(ref)}$ | $(0.6)$ | $0.17$ | $0.17$ | Volume fraction of the nucleus |
| $c_{ER}$ | – | – | $0.2$ | Volume fraction of the ER |
| $c_{CP}$ | $0.4$ | $0.83$ | $0.63$ | Volume fraction of the cytoplasm |

Abbreviations: EC = ellipsoidal cell; ER = endoplasmic reticulum; SC = spherical cell.

As suggested by Kotnik and Miklavčič [2000]; Peyman et al. [2007]; Merla et al. [2009], the dispersive properties of the EC, the CP and the NP were approximated using the dielectric functions of physiological salt solutions. To be more precise, they were based on measurements of a phosphate-buffered saline solution (PBS) at 27°C between 100 MHz and 2 GHz performed by Merla et al. [2009]. Based on an experimental assessment of the cytoplasmic conductivity in [Denzi et al., 2015], the conductivity of both the CP and NP was set to 0.32 S/m. Although the membranes of cells and organelles are complex (sub-)structures bound by proteins and filaments of the cytoskeleton, and thus are not a simple phospholipid bilayer (see Weaver [2003]; Gowrishankar et al. [2006]; Gowrishankar and Weaver [2006]), the material parameters of the membrane were assigned values based approximately on measurements of unilaminar liposome vesicles. These measurements were also performed in Merla et al. [2009]. The material coefficients corresponding to the compartments are summarized in Table 2 and are in accordance with measurements documented in scientific literature (see Ermolina et al. [2000]).

The relative permittivity and the conductivity of these compartments are displayed in Fig. 2a and b, respectively. The dispersive permittivities of the EC, the CP, and NP are identical, as are those of the PM and the nuclear envelope (NE). These spectral responses for the permittivities can be divided into three frequency intervals corresponding to the changes in the dielectric function of the material properties with frequency, i.e. the start and end of the roll-off of the two particular characteristic responses within the investigated range. These intervals are highlighted in shades of blue in Fig. 2a. The conductivity of the compartments can also be divided into three intervals, partially overlapping those of the permittivity, highlighted in shades of red at the top of Fig. 2b. The overlapping ranges of these intervals lead to further segmentation of the investigated spectrum into five frequency ranges highlighted in shades of beige which are enumerated by circled numbers in Fig. 2a.

**TABLE 2. Debye Coefficients of the Underlying Materials Related to the Specific Cell Components**

| Compartment | $\varepsilon_{r,s}$ | $\varepsilon_{r,\infty}$ | $f_0$ (GHz) | $\sigma_s$ (S/m) |
|---|---|---|---|---|
| Cytoplasm (CP) | 67 | 5 | 17.9 | 0.32 |
| Plasma membrane (PM) | 11.7 | 4 | 0.18 | 0.11e-6 |
| Nucleoplasm (NP) | 67 | 5 | 17.9 | 0.32 |
| Nuclear envelope (NE) | 11.7 | 4 | 0.18 | 0.11e-6 |
| Extracellular medium (EC) | 67 | 5 | 17.9 | 0.55 |

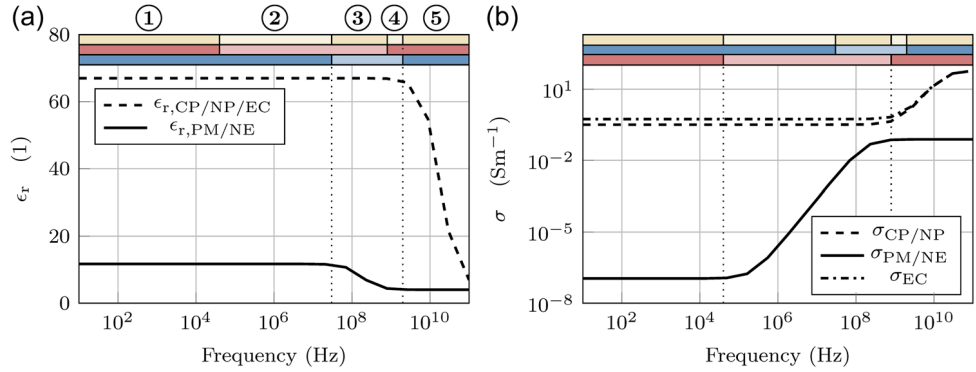

Fig. 2. Overview of the dispersive material parameters: (**a**) the relative permittivity of the cytoplasm (CP), nucleoplasm (NP), and the extracellular medium (EC) is represented as a dashed line, and that of the plasma membrane (PM) and the nuclear envelope (NE) as a solid line. (**b**) The conductivity of the CP and NP is represented as a dashed line, that of the PM and the NE as a solid line, and that of the EC as a dash-dotted line. The bar displayed in shades of blue corresponds to changes in the spectral response of the relative permittivity while the bar displayed in shades of red corresponds to changes in the conductivity. The overlapping ranges of these intervals lead to further segmentation of the investigated spectrum to five frequency ranges highlighted in shades of beige which are enumerated by circled numbers.

To provide a better overview, these ranges are summarized in Table 3. As outlined above, the frequency ranges defined by changes in the dispersive material properties of the cell compartments and the EC will also be plotted in shades of beige above any further graph given in the results section (Section Results) to show any correspondence to changes in the effective material properties and microdosimetric measures under investigation. These ranges will be referred to as material ranges (MR) 1-5.

**Simulation Setup**

The three cell models displayed in Fig. 1a–c were placed in a virtual parallel plate capacitor setup to conduct a quasi-static EM analysis between 10 Hz and 100 GHz. This capacitor setup was implemented in the FEM-based software package *COMSOL Multiphysics* [COMSOL, 2021]. The computational domain containing the centered spherical cell is shown as an illustrative example in Fig. 3a. This figure also shows the assignment of the dispersive material properties to the individual cell compartments. The edge length of the cube-shaped computational domain was 50 μm and was chosen to be large enough to minimize the influence of the depolarization fields around the cells on the neighboring cells. The resulting volume fraction of a cell in the computational domain was 1.15 % for the spherical cell and 4.12 % for the ellipsoidal ones. In the cross-section through the computational domain as shown in Fig. 3b, a time-harmonic voltage with constant amplitude, $U_0 = \varphi_1 - \varphi_0$, was applied between two opposing outer boundaries of the domain that were designed to function as electrodes (i.e. Dirichlet boundary conditions). $U_0$ was arbitrarily set to 1 V. The four remaining boundaries were defined with periodic boundary conditions (PBC) in order to suppress fringing fields and to reduce the memory resources of the subsequent quasi-static EM simulation. Due to this setup, the (computational) unit cell is effectively periodically extended in each direction indicated by the PBC. This approach was used in the past to investigate the effective material parameters of randomized (bio-) composites [Krakovsky and Myroshnychenko, 2002; Jerbic et al., 2020], and single cells [Huclova et al., 2010; Froehlich et al., 2014]. A description of how these simulations are verified by mixing rules is given in the Appendix. In order to cope with the anisotropy of the (ellipsoidal) cells under investigation, the voltage $U_0$ is then applied in the other (orthogonal) directions by swapping the electrodes with the PBCs (and vice versa). The low volume fraction in combination with this setup (i.e. the electrodes and PBCs) allows the interpretation of the cell under investigation either as a representative of a cell suspension corresponding to the static mean [Ermolina et al., 2000], or as a specific single cell [Chien et al., 2018; Guo et al., 2021].

**TABLE 3. Frequency Subintervals (i.e. material ranges [MR]) Corresponding to Changes in the Dispersive Material Parameters of the Cell Compartments and the Extracellular Medium Displayed in Figure 2 Highlighted in Shades of Beige at the Top of the Graphs**

|  | MR 1 | MR 2 | MR 3 | MR 4 | MR 5 |
|---|---|---|---|---|---|
| $f$ | 10–40 kHz | 40 kHz to 30 MHz | 30–800 MHz | 800 MHz to 2 GHz | 2–100 GHz |

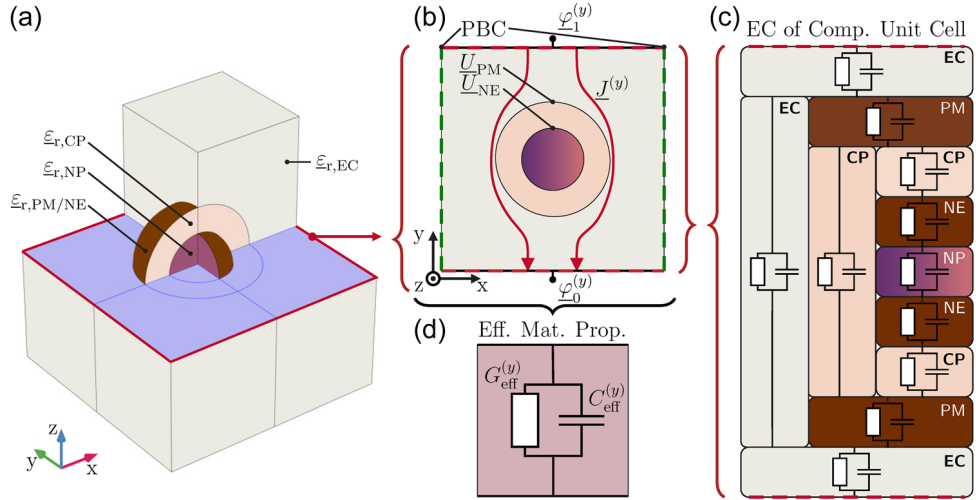

Fig. 3. Overview of the virtual capacitor experiment used for the quasi-static EM analysis of the computational domain: **(a)** Computational domain containing the spherical cell model as an illustrative example showing the assignment of the dispersive material properties introduced in Section "Derivation of the Dispersive Material Properties Based on Published Data" to the individual cell compartments. **(b)** Cross-section through the computational domain. The voltage, $U_0 = \varphi_1 - \varphi_0$, was applied to two opposing sides of the virtual capacitor set up which were designed to function as electrodes. Periodic boundary conditions (PBC) were applied to the remaining sides of the computational domain. This set up can be interpreted as a spatially periodic (computational) unit cell. To determine the voltages induced across both the plasma membrane (PM), $U_{PM}$, and the nuclear envelope, $U_{NE}$, their values were monitored at the closest points of the membranes facing the electrodes. **(c)** Equivalent circuit representation of the computational domain illustrating all the possible circuit paths summarized by RC shunt impedances. Using this equivalent circuit model, it is possible to make qualitative statements about how changes in the geometry of the cell affect the shunt elements in the individual current paths. In combination with the material ranges introduced in Section "Investigations to Determine the Necessity to Consider Membranes in Cell Modeling using the Simple Spherical Cell Model", the conceptual perception of the computational domain as an electrical network allows the qualitative separation and distinction of the influence of dispersive material properties and changes in the geometric structure of the cell. A detailed description of this concept is given in Section "Investigations to Determine the Necessity to Consider Membranes in Cell Modeling using the Simple Spherical Cell Model". The capitals in each circuit path correspond to the individual compartments of the unit cell which are the extracellular medium (EC), and those within the cell, namely, the PM, the cytoplasm (CP), the nuclear envelope (NE), and the nucleoplasm (NP). **(d)** Numerical homogenization summarizing the computational domain as an overall shunt impedance allowing the calculation of its effective macroscopic material parameters as outlined in Section "Investigations to Determine the Necessity to Consider Membranes in Cell Modeling using the Simple Spherical Cell Model".

In order to avoid high aspect ratios in the modeling mesh, the PM and the membranes of the cell organelles are modeled by a set of equations taking into account the conduction and the displacement current

$$\vec{n} \cdot \vec{\underline{J}}_{up} = \frac{1}{t_M}(\sigma_M + j\omega\varepsilon_0\varepsilon_{r,M})(\underline{\varphi}_{up} - \underline{\varphi}_{down}) \quad (2)$$

$$\vec{n} \cdot \vec{\underline{J}}_{down} = \frac{1}{t_M}(\sigma_M + j\omega\varepsilon_0\varepsilon_{r,M})(\underline{\varphi}_{down} - \underline{\varphi}_{up}) \quad (3)$$

where $\varphi_{up/down}$ and $\overrightarrow{n} \cdot \overrightarrow{J}_{up/down}$ denotes the membrane potentials and current densities normal to the upper and lower surfaces of the membrane. $t_M$ represents the thickness of the membrane while $\sigma_M$ and $\varepsilon_{r,M}$ represent the conductivity and the relative permittivity of the membrane respectively. $\varepsilon_0$ denotes the vacuum permittivity. Referring to Equations (2) and (3), thin membranes can now be represented by these tailored boundary conditions yielding a highly economic treatment as already proposed in Huclova [2011].

The simulation was performed on a PC equipped with two Intel Xeon E5-2697 V4 processors (36 cores) and 512 GB of RAM. In this study, a frequency range between 10 Hz and 100 GHz was investigated and sampled with 50 frequency points distributed logarithmically. The data points in between these sampled points were interpolated using piecewise cubic Hermite splines in postprocessing. The simulation for each frequency point lasted approximately 30 min using a MUMPS solver. The postprocessing of the data was performed within the MATLAB programming environment (version R2020a) [MATLAB, 2020].

## Conducted Investigations

Based on the concepts and models introduced above, the investigations conduced will be outlined in detail in the following. These investigations had two goals: to determine (i) the influence of internal organelles on the field distribution within a cell (i.e. microdosimetry) and (ii) the effect of this influence on the effective material properties of the simulated cell (suspension).

**Investigations to Determine the Necessity to Consider Membranes in Cell Modeling using the Simple Spherical Cell Model.** The field distribution within the spherical cell model (see Fig. 3a) is studied in order to investigate whether cell membranes must be considered over the entire investigated frequency range (i.e. 10 Hz to 100 GHz) or whether they may be neglected since their capacitive nature causes them to be bypassed at high frequencies. In these investigations, two aspects are of primary interest: (i) The electric potential and the resulting losses occurring at both the PM and the nuclear envelope under the application of an electric field and (ii) the effect of these membranes on the effective material properties of the simulated cell (suspension). The first aspect was addressed by adopting the methodological approaches used by Kotnik and Miklavčič [2000]; Kotnik and Miklavčič [2006] and the second aspect was considered by following [Ermolina et al., 2000; Feldman et al., 2003]. However, in contrast to the cited

studies where the problems were solved analytically, here, numerical methods were employed allowing the verification of the implemented computational model for the case of a spherical cell via comparison with the analytical results from the references listed above. In the following, we will outline the quantities involved in this numerical investigation.

To take the first aspect into consideration, the induced potential difference across the PM, $U_{PM}$, and that across the nuclear envelope, $U_{NE}$, were monitored as depicted in Fig. 3b. The volumetric loss densities corresponding to the induced membrane voltages can be calculated using $p_i = \sigma_i \cdot (U_i/t_i)^2$, expressed in W/m³, where $t_i$ and $\sigma_i$ represents the thickness and conductivity of the membrane, respectively, and the index $i \in \{PM, NE\}$ denotes either the PM or the nuclear envelop. However, the arbitrary choice of the voltage applied to the computational domain, $U_0$, means that it is relative rather than absolute values that are significant. The relative values are calculated for the membrane voltages, $U_i$, by normalizing to $U_0$ and for the loss densities, $p_i$, by normalizing to the maximum observed loss density, $p_{max}$. $p_{max} = \max(\max_f p_{PM}(f), \max_f p_{NE}(f))$ and is defined as the maximum loss density detected over the investigated frequency interval at either the PM or the nuclear envelope.

Moving on to the second aspect requiring consideration, the effective material parameters of the cell (suspension) were determined using the virtual capacitor experiment as depicted in Fig. 3d. The time-harmonic quasi-static EM analysis in the form of the capacitor setup presented in Section "Simulation Setup" leads to an effective admittance that is represented by an equivalent electrical parallel circuit consisting of the elements $G^{eff}(\omega)$ and $C^{eff}(\omega)$. This is given by

$$Y^{eff}(\omega) = \frac{I(\omega)}{U_0} = G^{eff}(\omega) + j\omega C^{eff}(\omega) \quad (4)$$

where the applied voltage, $U_0$, and the resulting current, $I(\omega)$, are directly accessible via COMSOL Multiphysics for an angular frequency $\omega = 2\pi f$. The dispersive effective material properties, $\varepsilon_r^{eff}(\omega)$ and $\sigma^{eff}(\omega)$, are thus easily deduced according to

$$Y^{eff}(\omega)\frac{d}{A} = \frac{I(\omega)}{U_0}\frac{d}{A} = \underbrace{\sigma^{eff}(\omega) + j\omega\varepsilon_0\varepsilon_r^{eff}(\omega)}_{\sigma^{eff}(\omega)} \quad (5)$$

where $d$ is the parallel plate distance and $A$ is the area of the electrode. In Equation (5), the right-hand term can be interpreted as the complex effective conductivity of the homogenized effective material, $\underline{\sigma}^{\text{eff}}(\omega)$, from which the required material parameters directly follow:

$$\sigma^{\text{eff}}(\omega) = \text{Re}\{\underline{\sigma}^{\text{eff}}(\omega)\} \qquad (6)$$

$$\varepsilon_r^{\text{eff}}(\omega) = \frac{\text{Im}\{\underline{\sigma}^{\text{eff}}(\omega)\}}{\omega\varepsilon_0} \qquad (7)$$

In order to consider anisotropies in the effective material, the quasi-static capacitor analysis is performed with excitation in the x, y, and z directions yielding corresponding frequency-dependent second-rank tensors for the effective conductivity,

$$\overleftrightarrow{\sigma}^{\text{eff}}(\omega) = \begin{pmatrix} \sigma_x^{\text{eff}}(\omega) & 0 & 0 \\ 0 & \sigma_y^{\text{eff}}(\omega) & 0 \\ 0 & 0 & \sigma_z^{\text{eff}}(\omega) \end{pmatrix} \qquad (8)$$

and for the effective permittivity,

$$\overleftrightarrow{\varepsilon}_r^{\text{eff}}(\omega) = \begin{pmatrix} \varepsilon_{r,x}^{\text{eff}}(\omega) & 0 & 0 \\ 0 & \varepsilon_{r,y}^{\text{eff}}(\omega) & 0 \\ 0 & 0 & \varepsilon_{r,z}^{\text{eff}}(\omega) \end{pmatrix}. \qquad (9)$$

As originally discussed in Section "Eukaryotic Cell Models and Spatial Segmentation", the numerical quantities addressing both aspect one and aspect two will be investigated in the context of a parameter analysis that varies the diameter of the nucleus so that the share of the cell volume, $c_N$, covers 0.3 to 0.9.

The addition of variable geometry, in the form of $c_N$, alongside the dispersive material properties of the individual cell compartments requires additional analysis to that found in the previously cited studies, which considered only constant material parameters. This analysis is needed to identify variations in cell geometry (i.e. in $c_N$), rather than the dispersive material properties of the individual cell components, as being the cause of dispersive characteristics in the spectral analysis of the effects under investigation. Given this requirement, the methodologies adopted from the cited studies need to be conceptually extended by considering the computational domain as an electrical network. An example of this is shown in Fig. 3c where an equivalent circuit model of the simple spherical cell is given in which all the possible current paths are summarized by RC shunt impedances. The abbreviations EC, PM, CP, NE, and NP denote the ellipsoidal cell, the plasma membrane, the cytoplasm, the nuclear envelope, and the nucleoplasm, respectively. Using this equivalent circuit model, it is possible to make qualitative statements about how changes in the geometry of the cell effect the shunt elements in the individual current paths. Increasing the nuclear radius, for example, results in a lower shunt conductance in the cytoplasmic pathway. These considerations, in combination with the MR defined in the previous subsection, provide a framework to qualitatively separate and distinguish the influence of dispersive material properties and changes in the geometric structure of the cell. In practice, this is performed by plotting the MR above the spectral responses of each quantity under investigation. The ranges then allow the changes in the underlying material properties to be related to changes in the dispersive characteristics being observed. The implementation of this concept can be seen in all of the illustrations shown in the results section.

**Investigations to Determine the Influence of Organelles on the Field Distribution in the Interior of Eukaryotic Cells and on Effective Macroscopic Material Properties.** In order to investigate the influence of organelles enclosed by a membrane, the field distribution within the cell in both ellipsoidal models is studied. In these investigations, two aspects are of primary interest: (1) The *absolute field distribution* within the cell model *with an ER* and (2) the *differences in the field distribution* comparing the models *with and without an ER*. These aspects will be addressed in a threefold analysis. First, the field distribution of current and volumetric loss density within the ellipsoidal cell models will be graphically analyzed in the symmetry plane which is parallel to the electrodes. After that, spatial minima and maxima will be determined and analyzed in the investigated frequency range (10 Hz to 100 GHz). Finally, the fields will be integrated using the spatial segmentation introduced earlier providing an overview of the currents flowing through and losses occurring within the cell models.

In order to introduce this field analysis in detail, the distribution of the absolute magnitude of the current density, $|\vec{J}|$, in A/m$^2$ and the volumetric loss density, $p$, in W/m$^3$ are plotted in Fig. 4b for the cell models with and without an ER. As illustrated in Fig. 4a, the field distributions of both quantities are displayed in the XZ plane under application of the voltage $U_0$ in the y-direction at 10 Hz. According to the color scale on the right of Fig. 4b, red areas indicate maximum values and light areas indicate minimum values.

Considering $|\vec{J}|$ and $p$ as a function of space, $\vec{r}$, and frequency, $f$, the comparison of the field distributions of both ellipsoidal models yields the spatial differences of the current density, $\Delta J$, and the volumetric loss density, $\Delta p$, which are shown in Fig. 4c and defined as follows:

$$\Delta J(f, \vec{r}) = \frac{|\vec{J}(f, \vec{r})|_{\text{ER}} - |\vec{J}(f, \vec{r})|_{\text{EC}}}{\langle|\vec{J}(f, \vec{r})|\rangle_{\text{ER}}} \quad (10)$$

$$\Delta p(f, \vec{r}) = \frac{p(f, \vec{r})_{\text{ER}} - p(f, \vec{r})_{\text{EC}}}{\langle p(f, \vec{r})\rangle_{\text{ER}}} \quad (11)$$

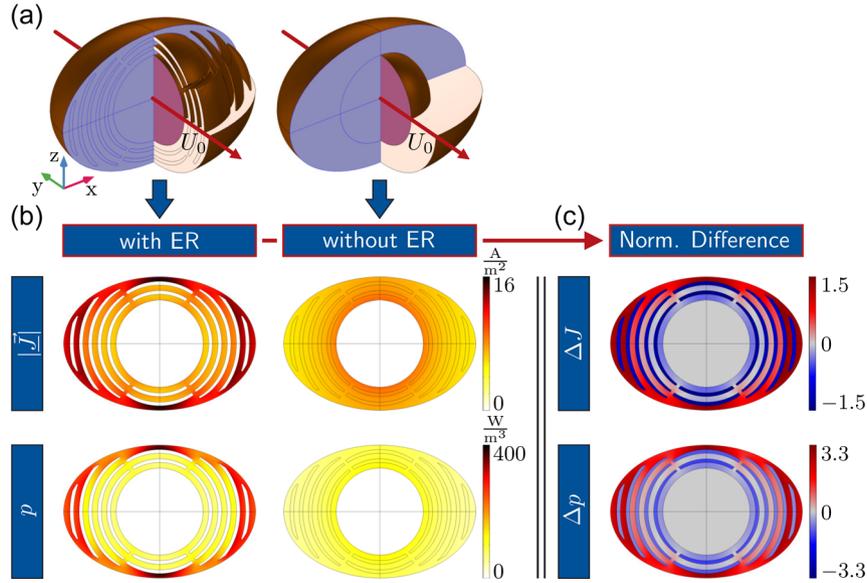

Fig. 4. Illustrative example of the stages of the field analysis using the ellipsoidal models at 10 Hz: **(a)** Ellipsoidal cell models highlighting the plane of investigation (XZ plane) in blue. The red arrows passing through both ellipsoidal cell models indicate an applied voltage $U_0$ in the $y$-direction. **(b)** The current density, $|\vec{J}|$, and the volumetric loss density distribution, $p$, are displayed in the XZ plane at 10 Hz. Maximum values are indicated by red areas and minimum values by light areas. The field plots can be observed while applying a voltage in the $y$-direction. **(c)** The distribution of the differences in current density, $\Delta J$, and the volumetric loss density, $\Delta p$, displayed in the XZ plane at 10 Hz. $\Delta J$ and $\Delta p$ are defined as formulated in Equations (10) and (11). Maximum values are indicated by red areas and minimum values by blue areas. The field plots can be observed while applying a voltage in the $y$-direction.

The indices ER and EC denote the ER and the ellipsoidal cell respectively, and $\langle|\vec{J}(f, \vec{r})|\rangle_{\text{ER}}$ and $\langle p(f, \vec{r})\rangle_{\text{ER}}$ are the spatial averages within the whole cell *with* an ER. Thus, $\Delta J$ and $\Delta p$ represent the spatial differences within the cell models normalized to the spatial average as a function of frequency and position. This is graphically illustrated by the red boxes around the column headers and the signs between them in Fig. 4. From the field plots shown in Fig. 4b, it can be seen that some regions within the cell model with an ER have higher current and volumetric loss densities than the same regions in the

cell model without an ER, and vice versa. Therefore, the visualization of these differences is adjusted by using a symmetrized color scale in Fig. 4c. According to the numerator terms in Equations (10) and (11), the red (positive) and blue (negative) areas in Fig. 4c can be interpreted as follows. In the red areas, the *model with an ER* has field strengths that are higher than those of the *model without an ER*, and in the blue areas, the *model with an ER* has field strengths that are lower than those of the *model without an ER*. This graphical evaluation shows in which areas the field magnitudes are overestimated or underestimated in a model *without* organelles. Grey regions indicate areas where there are no, or only very small, differences between the models with and without an ER.

Moving away from this illustrative example investigating a single frequency towards the general broadband investigation conducted in this study, it is necessary to interrelate and quantitatively compare the spatial minimum and maximum field values over the entire spectrum. For the current density, $|\vec{J}|$, and the volumetric loss density, $p$, this is done by determining the spatial minimum and maximum values of $J_{\text{min/max}}$ and $p_{\text{min/max}}$ over the entire frequency range investigated:

$$J_{\min/\max}(f) = \begin{cases} \min_{\vec{r}} \left| \vec{J}(f, \vec{r}) \right| \\ \max_{\vec{r}} \left| \vec{J}(f, \vec{r}) \right| \end{cases} \qquad (12)$$

$$p_{\min/\max}(f) = \begin{cases} \min_{\vec{r}} p(f, \vec{r}) \\ \max_{\vec{r}} p(f, \vec{r}) \end{cases} \qquad (13)$$

However, due to the arbitrary choice of the voltage $U_0$, only relative values are of significance, and thus, $J_{\min/\max}$ and $p_{\min/\max}$ need to be normalized to the global maxima observed in the entire frequency range. Defining $J_{\mathrm{abs,max}}$ and $p_{\mathrm{abs,max}}$ as

$$J_{\mathrm{abs,max}} = \max_f J_{\max}(f) \qquad (14)$$

and

$$p_{\mathrm{abs,max}} = \max_f p_{\max}(f), \qquad (15)$$

the normalized logarithmic amplitude ratios

$$J_{\mathrm{n,min/max}}(f) = 20 \, \log_{10}\left( J_{\min/\max}(f) \Big/ J_{\mathrm{abs,max}} \right) \qquad (16)$$

and

$$p_{\mathrm{n,min/max}}(f) = 10 \, \log_{10}\left( p_{\min/\max}(f) \Big/ p_{\mathrm{abs,max}} \right). \qquad (17)$$

can be calculated and used for spectral analysis.

In contrast, for the spatial differences of the current density, $\Delta J$, and the loss density, $\Delta p$, a spectral analysis can be easily performed by determining the respective minima and maxima over the frequency range:

$$\Delta J_{\min/\max}(f) = \begin{cases} \min_{\vec{r}} \Delta \vec{J}(f, \vec{r}) \\ \max_{\vec{r}} \Delta \vec{J}(f, \vec{r}) \end{cases} \qquad (18)$$

$$\Delta p_{\min/\max}(f) = \begin{cases} \min_{\vec{r}} \Delta p(f, \vec{r}) \\ \max_{\vec{r}} \Delta p(f, \vec{r}) \end{cases} \qquad (19)$$

This is possible since $\Delta J$ and $\Delta p$ represent comparative quantities between two models which are already normalized to the spatial average of the respective quantities occurring in the ellipsoidal cell model with an ER.

To complete the threefold analysis, the focus now turns to the relative current and loss distributions within the cell rather than the spatial distributions of current and volumetric loss densities. These relative quantities are defined as

$$\frac{I_i}{I_{\mathrm{cell}}} \qquad (20)$$

$$\frac{P_i}{P_{\mathrm{cell}}} \qquad (21)$$

where $i \in \{\mathrm{CP; NP; ER}\}$ denotes the cytoplasm, nucleoplasm, or endoplasmic reticulum. The currents in each compartment,

$$I_i(f) = \iint \vec{J}(\vec{r}, f) \cdot \vec{n} \; dA_i, \qquad (22)$$

are obtained by integrating the current densities in the sectional planes of the cell under investigation (see Fig. 1g–i) over the surface area of each compartment (see. Fig. 1j and k) with $\vec{n} \in \{\vec{e}_x, \vec{e}_y, \vec{e}_z\}$. Correspondingly, the losses in each compartment,

$$P_i(f) = \iiint p(\vec{r}, f) dV_i, \qquad (23)$$

are obtained by integrating the loss densities over the volume of the individual compartments (see Fig. 1d–f). The cell currents and cell losses are defined as

$$I_{\mathrm{cell}}(f) = I_{\mathrm{CP}}(f) + I_{\mathrm{NP}}(f) + I_{\mathrm{ER}}(f) \qquad (24)$$

$$P_{\mathrm{cell}}(f) = P_{\mathrm{CP}}(f) + P_{\mathrm{NP}}(f) + P_{\mathrm{ER}}(f). \qquad (25)$$

In addition to the microdosimetric analysis presented above, the effect of a modeled ER on the effective macroscopic material properties will be assessed by applying the approach outlined in Section "Investigations to Determine the Necessity to Consider Membranes in Cell Modeling using the Simple Spherical Cell Model".

## RESULTS

The results of the investigations to determine the need for membrane modeling are presented in Section "The Necessity to Consider Membranes in Cell Modeling using the Spherical Cell Model" using the simple spherical cell model. As already described in detail in Section "Investigations to Determine the Necessity to Consider Membranes in Cell Modeling using the Simple Spherical Cell Model", this is done in the context of a parameter analysis varying the diameter of the nucleus ($c_N \in [0.3, 0.9]$) along with a broadband frequency analysis taking account of dispersive material properties. The frequency ranges corresponding to changes in the underlying material properties (i.e. MR) are plotted above the spectral responses of each quantity under investigation and help to assign observed effects either to changes in the material parameters or to effects caused by the geometrical composition of the cell model. The MR also appear above the results of the investigations examining the influence of organelles enclosed by a membrane on the field distribution inside the cell as presented in Sec. 3.2.

## The Necessity to Consider Membranes in Cell Modeling using the Spherical Cell Model

**Induced membrane voltage and the resulting losses occurring at both the plasma membrane and the nuclear envelope.** In Fig. 5a, the induced membrane voltages normalized to the voltage applied to the computational domain, $U_0$, can be observed. The blue curves correspond to the voltage monitored at the PM, $U_{PM}$, and the red curves correspond to the voltage monitored at the nuclear envelope, $U_{NE}$, (see. Fig. 3b). The dashed, solid, and dashed-dotted lines show the results of cell models with a volume fraction of the nucleus in the cell of $c_N = 0.3$, $c_N = 0.6$, and $c_N = 0.9$, respectively.

In general, $U_{PM}$ shows low-pass behavior and $U_{NE}$ shows bandpass behavior. The parameter analysis leads to the identification of two effects on the induced membrane voltage: (i) The larger the nuclear radius, the lower the frequency at which the transitions to low-pass behavior across the PM can be observed. (ii) Correspondingly, bandpass behavior also shifts to lower frequencies while also demonstrating higher induced voltages at the nuclear envelope. The range of observed membrane voltages corresponding to the $c_N$ values under investigation is confined by the dotted (i.e. $c_{cell} = 0.3$) and dashed-dotted (i.e. $c_{cell} = 0.9$) lines and highlighted by the gray-shaded areas in between.

To separate the influence of dispersive material properties from that of changes in cell geometry, the observed effects on membrane voltages are evaluated with respect to the MR represented by the beige bars above Fig. 5a. The initial changes in the response of the membrane voltages induced at both the nuclear envelope and the PM occur in MR 1 and thus, are not caused by any change in the underlying material parameters. Consequently, the shift of the onset of the observed effects can be explained by the geometrical composition of the cell model. Considering the equivalent circuit model of the computational domain in Fig. 3c, two parallel paths exist within the cell, one leading through the nucleus and one bypassing it. In the case of a larger nucleus, the cross-sectional area of the bypassing CP path is reduced resulting in lower conductance. This, in turn, leads to a higher impedance of the whole cell interior (i.e. CP and nucleus path) causing an significant increase of the induced membrane voltage at the nuclear envelope at lower frequencies. At frequencies larger than 5 MHz in MR 2, the conductivity of the membrane increases and reduces the difference between the impedances of the circuit passing through the nucleus and the one bypassing it. As a result, the voltages $U_{PM}$ and $U_{NE}$ of the different model variants, indexed by the individual volume fractions, $c_N$, begin to align and seem to converge to a single curve again.

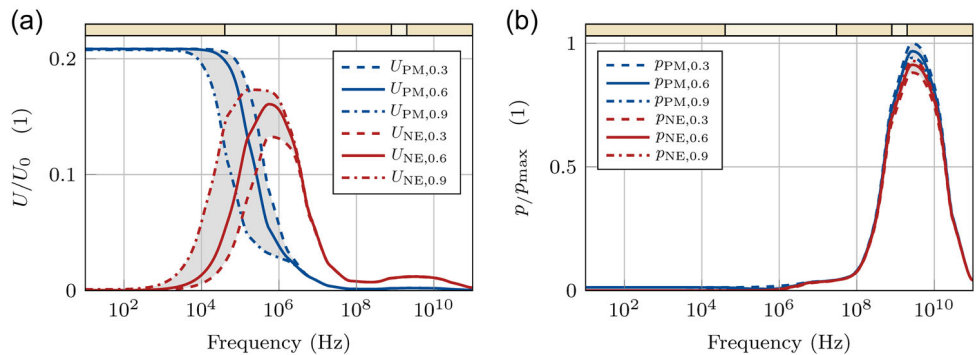

Fig. 5. (**a**) Overview of the induced membrane voltages normalized to the voltage applied to the computational domain, $U_0$ and (**b**) the volumetric loss density normalized to the maximum observed volumetric density, $p_{max}$. The blue curves show these measures at the plasma membrane (PM), the red ones at the nuclear envelope (NE). The dashed, solid, and dashed-dotted lines show the results of cell models with a volume fraction of the nucleus in the cell of $c_N = 0.3$, $c_N = 0.6$, and $c_N = 0.9$, respectively.

The volumetric loss density normalized to $p_{max}$ is displayed in Fig. 5b. The blue curves correspond to the loss densities at the PM, $p_{PM}$, and the red curves correspond to the corresponding position on the nuclear envelope, $p_{NE}$. The dashed, solid, and dashed-dotted lines show the results of cell models with a volume fraction of the nucleus in the cell of $c_N = 0.3$, $c_N = 0.6$, and $c_N = 0.9$, respectively.

Here it can be observed that for frequencies lower than 1 MHz, the volumetric loss density is relatively low compared to that at larger frequencies. This can be explained by the low conductivity of the membrane resulting in a barrier for every current penetrating it. For larger frequencies, the loss densities in both the PM and the nuclear envelope increase strongly, reaching a global maximum between 1 and 10 GHz. This increase is initiated due to the decreasing relative permittivity of the membranes in MR 3. The rapid decrease of the volumetric loss density after reaching the maximum is initiated by the decrease of the permittivity of the CP and the NP in MR 5. In contrast to the observed effects on the induced membrane voltages, the size of the nucleus has only little influence on the volumetric loss density. This is illustrated by the range of observed values confined by the dotted (i.e. $c_{cell} = 0.3$) and dashed (i.e. $c_{cell} = 0.9$) lines and highlighted by the gray-shaded areas in between.

**The Effect of Cell Membranes on the Effective Material Parameters of the Spherical Cell Model.** The effective material parameters of the simulated cell (suspension) are illustrated in Fig. 6a and b, showing the relative permittivity and the conductivity, respectively. The dashed, solid, and dashed-dotted lines show the results of cell models with a volume fraction of the nucleus in the cell of $c_N = 0.3$, $c_N = 0.6$, and $c_N = 0.9$, respectively.

The effective permittivity depicted in Fig. 6a clearly displays a range of values between 10 kHz and 1 MHz for the various models under investigation. It can be seen that the cell model with the larger nucleus leads to a reduced ability to accumulate charges, due to the previously discussed shift of the induced membrane voltages within the cell, especially at the PM. The size of the nucleus thus leads to a lowering of the characteristic frequency of the interfacial polarization. Due to the high conductivity of the EC, no variations within the effective conductivity of the suspension can be observed.

The cell models investigated were compared to analytical models based on mixing rules. The comparison showed maximum relative errors of <0.5 % for both conductivity and permittivity over the entire interval. The calculation rules for verification can be found in the Appendix.

**The Influence of Organelles on the Field Distribution in the Interior of Eukaryotic Cells and on Effective Macroscopic Material Properties**

Based on the threefold field analysis introduced in Section "Investigations to Determine the Influence of Organelles on the Field Distribution in the Interior of Eukaryotic Cells and on Effective Macroscopic Material Properties", the influence of organelles enclosed by a membrane on the field distribution inside the cell is investigated. The investigation begins with an analysis of the *absolute magnitude* of the current density, $|\vec{J}|$, and *volumetric* loss density, $p$, throughout the interior of the ellipsoidal cell model

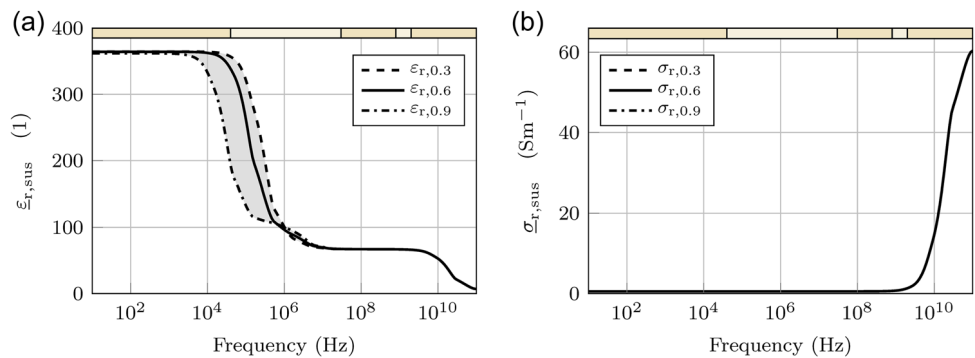

Fig. 6. Overview of the effective macroscopic material properties: (**a**) effective relative permittivity and (**b**) effective conductivity for volume fractions of the nucleus in the cell of $c_N = 0.3$, $c_N = 0.6$, and $c_N = 0.9$, respectively.

*with* an ER. Subsequently, the field distribution within the model with an ER is compared with that of the ellipsoidal cell model without an ER, focusing on the resulting differences that would arise in the modeling process if the cell interior were (over)simplified. Finally, the effect of an internal distributed membrane system on the effective macroscopic material properties of the cell is evaluated.

**Field Analysis of the Ellipsoidal Cell Model with an ER.** An overview of the field analysis investigating the influence of organelles enclosed by a membrane to the field distribution of the cell interior is provided by Fig. 7. In Fig. 7a, the current density, $|\vec{J}|$, and the volumetric loss distribution, $p$, are displayed in the XZ plane at different frequencies. Maximum values in each plot are indicated by red areas and minimum values by light areas. The field plots show data under the application of a voltage in the y-direction. To interrelate and quantitatively assess the minimum and maximum values of the color scale in the field plots in Fig. 7a, $J_{n,min/max}(f)$ and $p_{n,min/max}(f)$ are displayed in Fig. 7b and c, respectively.

In Fig. 7a, it can be seen that the cisternal organelle structures, which are concentrated around the nucleus, cause a heterogeneous field distribution in the CP. The layered structure of the membranes of the ER around the nucleus throughout the CP means that the current entering the cell is primarily conducted via the outer cell edge and the current and volumetric loss densities around and in the nucleus are comparatively low, especially at low frequencies. At a frequency of 10 Hz for example, it can be seen that almost no current flows or losses occur in the ER or the nucleus. However, these compartments become increasingly penetrable as the frequency increases. At frequencies of 10 MHz and above, a current distribution is established within the cell that corresponds approximately to the area of the respective compartments on the (cross-)sectional plane of the cell. In Fig. 7b, it can be observed that the maximum current density at 10 Hz is about 100 dB smaller than at 100 GHz. The reason for this is the high conductivity of the EC and the low conductivity of the PM. Roughly speaking, the current density within the cell increases strongly between 100 Hz and 1 MHz. Furthermore, the difference between the minimum and maximum current density decreases from about 90 dB to <1.5 dB. For frequencies larger than 1 MHz, the current density increases further and reaches a plateau around 100 GHz. To summarize, it can be noted that low cell currents with high relative current density differences occur at frequencies <1 MHz while high currents with lower differences are observed at frequencies >1 MHz. These statements can also be made for the volumetric loss density shown in Fig. 7c.

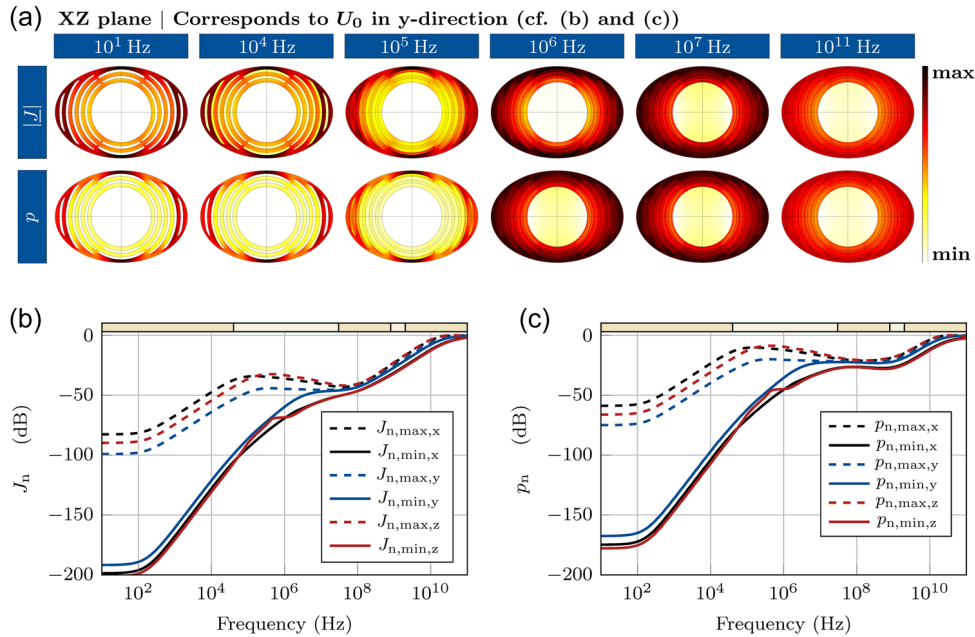

Fig. 7. Overview of the field analysis investigating the influence of organelles enclosed by a membrane on the field distribution of the cell interior: (**a**) The current density, $|\vec{J}|$, and the loss distribution, $p$, are displayed in the XZ plane at different frequencies. Maximum values are indicated by red areas and minimum values by light areas. The field plots can be observed while applying a voltage in the y-direction. (**b**, **c**). Quantitative assessment of the minimum and maximum values of both $J_n$ (see. Eq. (16)) and the $p_n$ (see. Eq. (17)) plotted on a logarithmic scale. The indices in the legend, x, y, and z, denote the direction of the voltage applied in the capacitor experiment.

Table 4 provides an overview of the normalized minimum and maximum current and volumetric loss densities in the frequency interval between 1 and 100 GHz. It is worth noting that the ratio between the minimum and maximum current density is between 1.016 and 1.157 and that of the volumetric loss density between 1.043 and 1.342, and thus, a nonnegligible difference between currents and losses in the outer edge of the cell interior and the nucleus exists, even at such high frequencies.

To be concise, a comprehensive field analysis of the cell model without an ER is omitted, since this information is inherently contained within the comparative field analysis between the two ellipsoidal models presented in the following subsection.

**Comparative Field Analysis of the Ellipsoidal Cell Models With and Without an endoplasmic reticulum.** An overview of the comparative field analysis investigating the influence of organelles enclosed by a membrane to the field distribution of the cell interior is provided by Fig. 8. Here, the emphasis is to contrast the differences in the field distribution when modeling the cell with and without an ER. In Fig. 8a, the distribution of the differences in the current density, $\Delta J$, and the volumetric loss density, $\Delta p$, is displayed in the XZ plane at different frequencies. $\Delta J$ and $\Delta p$ are defined as formulated in Eqs. (10) and (11) in Section "Investigations to Determine the Influence of Organelles on the Field Distribution in the Interior of Eukaryotic Cells and on Effective Macroscopic Material Properties". Maximum values in each plot are indicated by red areas and minimum values by blue areas. The field plots show data under the application of a voltage in the y-direction. To interrelate and quantitatively assess the minimum and maximum values of the field plots in Fig. 8a, $\Delta J_{min/max}(f)$ and $\Delta p_{min/max}(f)$ are displayed in Fig. 8b and c, respectively.

Observing the field distribution of $\Delta J$ and $\Delta p$ at low frequencies ($\leq$10 kHz) in Fig. 8a, it can be seen that an ER enclosed by a membrane has a large influence on the current and volumetric loss density, as these regions are electrically shielded. At frequencies larger than 10 kHz and less than 10 MHz, it becomes very clear that although shielding provided by the ER is not absolute, the current distribution is changed in such a way that currents are mainly conducted at the outer edge of the cell compared to the model without ER. In particular, the field images at 1 MHz show that the cascade-like concentric layering of the ER around the nucleus leads to its electrical shielding, thus, the ellipsoidal cell model without an ER overestimates the current flowing through or around the nucleus. Looking at the spectral responses of the minimum and maximum values of $\Delta J_{min/max}$ and $\Delta p_{min/max}$ illustrated in Fig. 8b and c, respectively, it can be seen that the differences between the current and loss densities in both models are relatively low for frequencies >10 MHz in the entire cell space.

Shifting attention from the analysis of the current and volumetric loss *density* distribution performed previously, the relative current and loss distribution within both cell models will now be analyzed using the integral quantities defined in Section "Investigations to Determine the Influence of Organelles on the Field Distribution in the Interior of Eukaryotic Cells and on Effective Macroscopic Material Properties" (see Eqs. (20) to (25)). In Fig. 9, the differences between the relative currents and losses in the individual cell compartments are contrasted in both ellipsoidal cell models. The currents and losses in the model *with an ER* are displayed in Fig. 9a and c, respectively, while Fig. 9b and d display these quantities in the model *without an ER*. In the legends, the indices CP, NP, and ER denote the CP, NP, and ER and x, y, and z, denote the direction of the voltage applied in the capacitor experiment.

**TABLE 4. Overview of the Normalized Minimum and Maximum Current and Loss Densities in the Frequency Interval Between 1 and 100 GHz**

Current density

| Frequency (GHz) | $J_{n,min,y}$ | $J_{n,max,y}$ | $J_{n,min,y} - J_{n,max,y}$ | $J_{min}/J_{max}$ |
|---|---|---|---|---|
| 100 | −0.182 | −0 | −0.182 | 1.016 |
| 50 | −1.012 | −0.87 | −0.142 | 1.021 |
| 10 | −7.908 | −6.74 | −1.168 | 1.144 |
| 1 | −26.9 | −25.63 | −1.27 | 1.157 |

**Volumetric loss density**

| Frequency (GHz) | $p_{n,min,y}$ | $p_{n,max,y}$ | $p_{n,min,y} - p_{n,max,y}$ | $p_{min}/p_{max}$ |
|---|---|---|---|---|
| 100 | −0.182 | −0 | −0.182 | 1.043 |
| 50 | −0.679 | −0.364 | −0.315 | 1.075 |
| 10 | −7.15 | −5.987 | −1.163 | 1.307 |
| 1 | −21.897 | −20.618 | −1.279 | 1.342 |

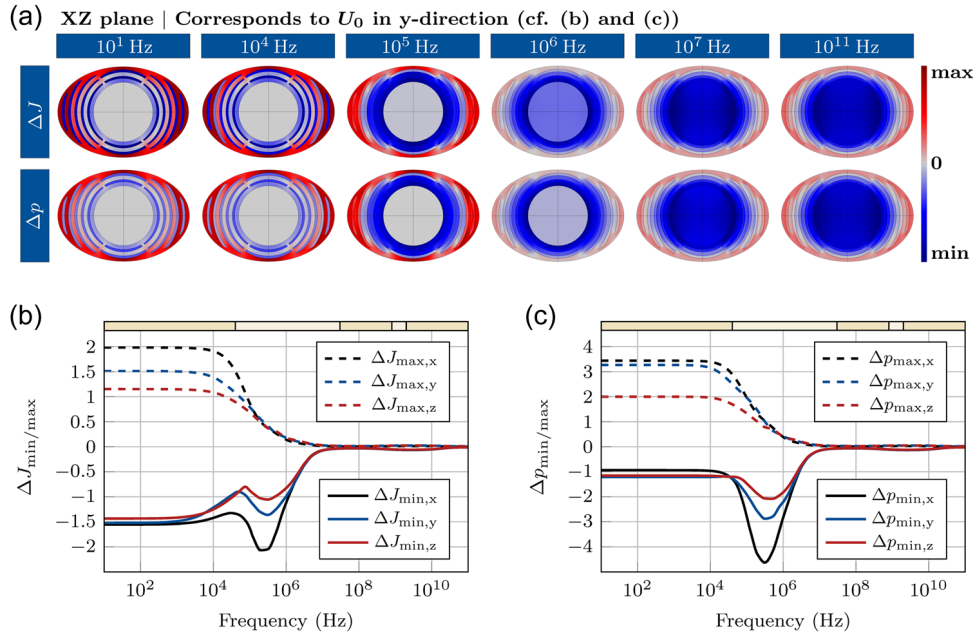

Fig. 8. Overview of the comparative field analysis investigating the influence of organelles enclosed by a membrane to the field distribution of the cell interior contrasting the differences occurring when modeling the cell with or without an ER: (**a**) The distribution of the differences in the current density, $\Delta J$, and the volumetric loss density, $\Delta p$, displayed in the XZ plane at different frequencies. $\Delta J$ and $\Delta p$ are defined as formulated in Eqs. (10) and (11). Maximum values are indicated by red areas and minimum values by blue areas. The field plots can be observed while applying a voltage in the y-direction. (**b**, **c**) Quantitative assessment of the minimum and maximum values of both $\Delta J$ (see Eq. (18)) and $\Delta p$ (see Eq. (19)) are plotted. The indices in the legend, x, y, and z, denote the direction of the voltage applied in the capacitor experiment. ER = endoplasmic reticulum.

In Fig. 9a for instance, it can be clearly seen that cell currents at frequencies <1 kHz are almost exclusively conducted in the CP, with little or no current in the ER and the nucleus. At frequencies >1 kHz, the cell current also flows through the cisternae of the ER. Furthermore, it can be seen that the nucleus only conducts a part of the cell current from a frequency >100 kHz upwards. At frequencies of 10 MHz and higher, a current distribution is established within the cell that corresponds approximately to the area of the respective compartments on the (cross-)sectional plane of the cell. Due to the different dimensions of the individual compartments in the (cross-)sectional plane, a slightly anisotropic current distribution can be observed. These observations can also be transferred to the relative cell losses displayed in Fig. 9c. Another finding is that the nucleus conducts current at higher frequencies than the rest of the organelles, because of the thicker membrane assumed in the model ($t_M = 7$ nm and $t_{NE} = 40$ nm). In addition, Fig. 9b and d display the relative current and loss distributions for the model without an ER to better relate them to those of the model with an ER.

**The Effect of an ER on the Effective Material Parameters of the Ellipsoidal Cell Models.** The effective material parameters of the simulated suspension are displayed in Fig. 10a and c for the cell model with an ER and in Fig. 10b and d for the model without an ER. The indices, x, y, and z, denote the direction of the voltage applied in the capacitor experiment, based on the semi-axes of the cell defined in these directions.

Comparing the effective macroscopic permittivity of both models, $\varepsilon_{r,sus}$, it can be observed that there is no difference in the static permittivity for frequencies <10 kHz. The direction of the applied voltage, however, leads to different values for the permittivity while the pattern of this anisotropy remains the same between the two models. In the frequency range highlighted in grey (i.e. between 10 kHz and 1 MHz), differences between the models can be observed. In the model with an ER, the characteristic frequency of the interfacial polarization process shifts towards lower frequencies, so that the observed Maxwell–Wagner roll-off starts at low frequencies. This is especially true when the voltage is applied in the *x*-direction, since in this case, the largest number of cisternae lies between the PM and the nuclear envelope (cf. Fig. 1c). This can be attributed to the changes in the transmembrane potential as discussed in earlier (see Section "The Effect of Cell Membranes on the Effective Material Parameters of the Spherical Cell Model"). At frequencies where the impedance of the PM loses its overriding dominance, the concentric arrangement of many cisternal membranes leads to a series of internal voltage drops and a potential distribution that is oriented toward the interior of the cell. For frequencies >1 MHz, all curves converge to the same value and show the same trends in both models. Due to the high conductivity of the EC, the ER has no impact on the effective conductivity of the suspension. As already observed for the spherical cell model, the effective conductivity of the ellipsoidal model is primarily governed by the material properties of the extracellular space.

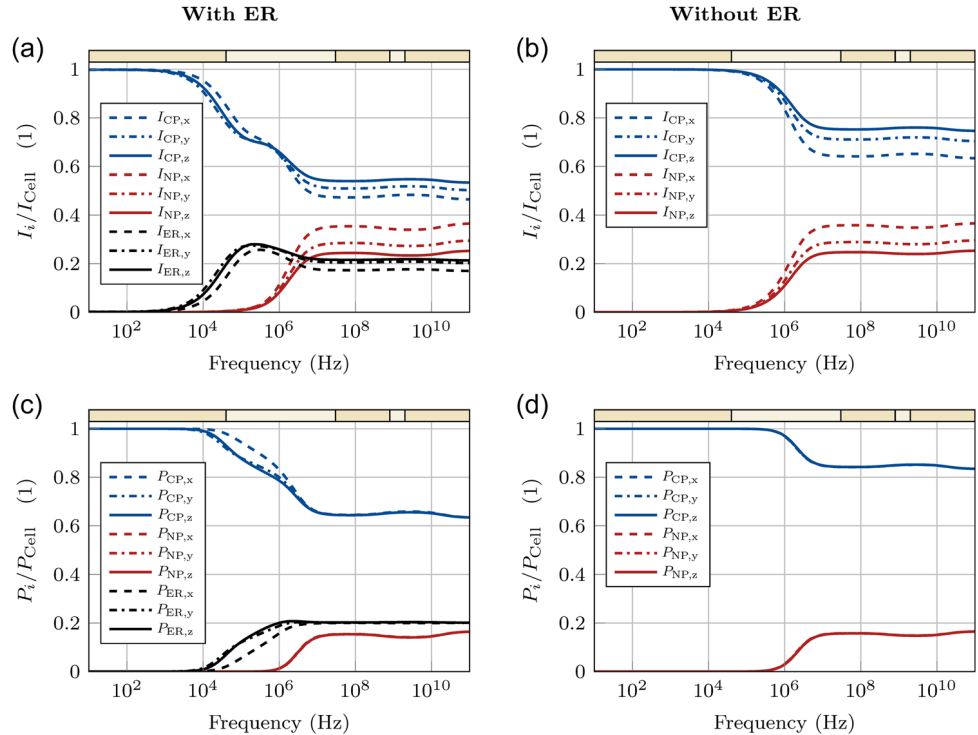

Fig. 9. Comparison of the spectral response of the relative current and loss distribution within the individual cell compartments between the cell model with an endoplasmic reticulum (ER) and the one without. (**a**, **c**) Display the relative currents and losses in the individual cell compartments in the ellipsoidal cell model with an ER, while (**b**, **d**) display these quantities in the model without an ER. The indices, CP, NP, and ER, denote the cytoplasm, nucleoplasm, and endoplasmic reticulum. The indices in the legend, *x*, *y*, and *z*, denote the direction of the voltage applied in the capacitor experiment.

## CONCLUSION

As part of a comparative study investigating the impact of increasing levels of detail in modeling eukaryotic cells, methods for graphical and quantitative analysis of the resulting field distribution within cell bodies have been presented. In addition, the qualitative separation of and distinction between the influence of dispersive material properties and changes in the geometric structure of the cell have been demonstrated.

Using a *simple spherical cell model*, it has been shown that the cell membrane has to be considered at low frequencies (<1 MHz) as well as at high frequencies (>1 MHz) for the following reasons. At low frequencies, the modeling of the membrane is necessary because, in the absence of induced physiological changes such as electroporation, the membrane electrically shields the interior of the cell and has a high impact on the effective relative permittivity of the cell system. At high frequencies, modeling of the membrane is necessary because the different characteristic frequencies of the dispersive material functions lead to peaks in the volumetric loss density in the membrane, which must be considered in the context of microdosimetric investigations. This conclusion is in agreement with analytical and numerical studies published in Kotnik and Miklavčič [2000] and Saviz and Faraji-Dana [2020].

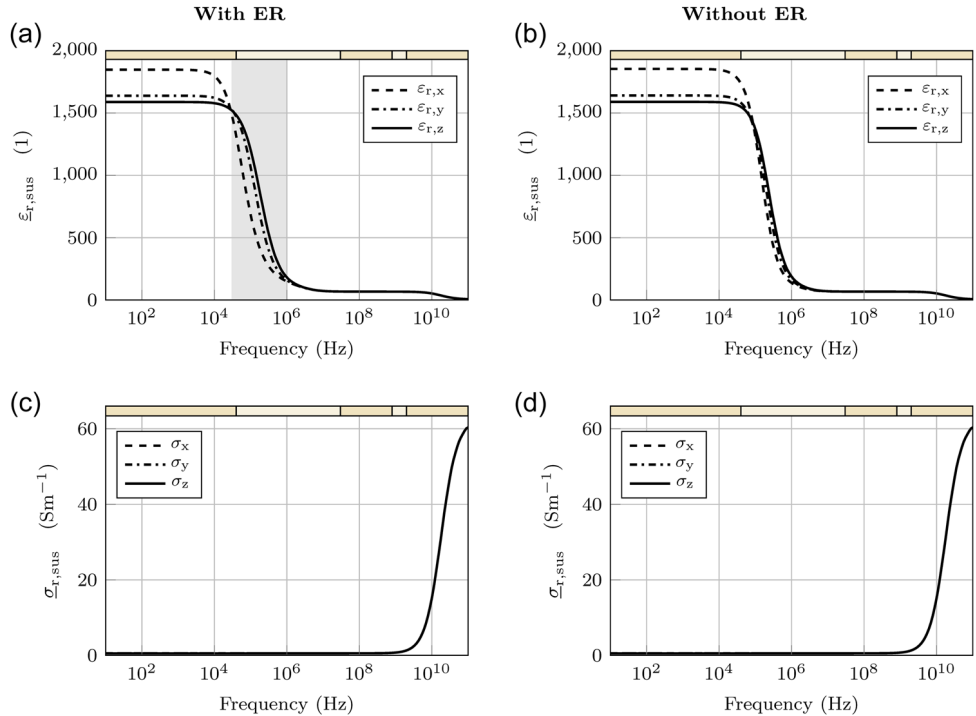

Fig. 10. Overview of the effective material properties of the ellipsoidal cell models: (**a**) and (**c**) display the effective permittivity, $\varepsilon_{r,sus}$, and conductivity, $\sigma_{sus}$, of the cell suspension with cells including an endoplasmic reticulum (ER), and (**b**, **d**) display the values of a suspension without an ER. The indices in the legend, *x*, *y*, and *z*, denote the direction of the voltage applied in the capacitor experiment.

Using an *ellipsoidal cell model involving an ER*, microdosimetric investigations have shown that a system of organelles enclosed by membranes has a large influence on the field distribution within the cell. At frequencies below 10 kHz, it has been demonstrated that organelle structures may act like barriers due to their material properties. As a result, cell currents tend to be conducted across the outer edge of the cell. In general, the maximum relative difference between the largest and smallest local current and loss densities occurs at such low frequencies, with the high resistance of the cell membrane allowing only small currents to enter the cell. At frequencies between 10 kHz and 10 MHz, the membranes become increasingly penetrable, causing the cell current to increase and the relative differences between minimum and maximum field strengths to decrease, resulting in the field distribution becoming increasingly homogeneous. Despite this trend toward a more homogeneous current and loss distribution, it has been demonstrated that the ratio between the maximum and the minimum current density is still 1.157 and that for the volumetric loss density is still 1.342 at frequencies as high as 1 GHz and should not be considered negligible.

Comparing *ellipsoidal cell models with and without an ER*, it has been shown that neglecting the cisternae of the organelles in the model without an ER leads to a strong underestimation of both the of the current and volumetric loss density distribution at the outer edge of the cell, while at the nucleus, both values are overestimated. This observation has been consistently made over the entire frequency range studied, despite the better conductivity of the membrane at higher frequencies. Thus, although the cisternae of the ER no longer act as rigid barriers, their membranes increase the resistance of the current paths that leads across the nucleus by layering many membranes on top of each other, as a result of which the current density in the nucleus is significantly higher in the model without an ER. In conclusion, microdosimetric studies evaluating potential health risks from EM exposure at the cellular level should therefore include cell organelles in modeling field-tissue interactions, even at frequencies used or being considered for use in 5 G telecommunications standards (600 MHz to 2.5 GHz and 24 GHz to 52 GHz).

## OUTLOOK

Using generic cell and organelle morphology, it has already been shown here that organelles such as the ER have a major influence on the internal current and loss distribution within a cell across the whole investigated frequency range. However, due to the generic modeling approach, the results of these studies can only be qualitatively applied to real cells. Therefore, to obtain more quantitative results, a useful next step would be to develop cell models that take into account the asymmetric, stochastic nature of real eukaryotic cells by first capturing the internal structure of as large a number as possible and transferring them to computer models. However, transferring realistic cell morphologies into simulation models is per se a major challenge, since, for example, the dimensions of organelle cisternae in reality are much smaller and the density of membrane layers much larger than in the cell models presented here. Another future step would be to simulate multicellular structures to determine the effects of different model details (e.g. the density of the ER) on the effective material properties of tissues in the context of multiscale modeling as developed in Froehlich et al. [2014]. This would allow the determination of the EM energy intake at the microstructural level in cells and their organelles as part of a realistic exposure scenario.

## ACKNOWLEDGMENTS

This work was funded by the Swiss Research Foundation for Electricity and Mobile Communications (FSM) at ETH Zürich in the framework of FSM-Project No. A2019-01. We also acknowledge the partial support from the Deutsche Forschungsgemeinschaft (DFG, German Research Foundation) under Project 287022738 of the CRC/TRR 196 MARIE in the context of project M03 and the partial funding by the project "Konzepte für Ende-zu-Ende Terahertz 6G-Mobilfunk" (no. 16KISKK039) in the framework of the BMBF (German Federal Ministry of Education and Research) joint research project 6GEM. Open Access funding enabled and organized by Projekt DEAL.

**APPENDIX**

In order to validate the computational model used to determine the effective material properties of the cell and cell suspension, the simulations considering the simple spherical cell model are compared with analytical mixing rules. As described in [Asami et al., 1989] and [Feldman et al., 2003], the effective permittivity of the cell $\varepsilon_{cell}$ is modeled as double-layered shell

$$\varepsilon_{cell} = \varepsilon_{PM} \frac{2(1-v_1) + (1+2v_1)E_1}{(2+v_1) + (1-v_1)E_1} \quad (26)$$

where the geometrical parameter $v_1$ represents for the volume ratio between the PM and the inner layers and the intermediate parameter $E_1$ the effective permittivity of the lower layers weighted by the permittivity of the upper layer, which is, in this case, the PM. The parameters $v_1$ and $E_1$ is given by

$$v_1 = \left(1 - \frac{t_M}{r_{cell}}\right)^3 \quad (27)$$

$$E_1 = \frac{\varepsilon_{CP}}{\varepsilon_{PM}} \cdot \frac{2(1-v_2) + (1+2v_2)E_2}{(2+v_2) + (1-v_2)E_2} \quad (28)$$

Similar to the calculation of the effective permittivity of the whole cell, the intermediate parameter $E_1$ itself is also calculated from the intermediate parameters $v_2$ and $E_2$. These can be calculated according to

$$v_2 = \left(\frac{r_N}{r_{cell} - t_M}\right)^3 \quad (29)$$

$$E_2 = \frac{\varepsilon_{NP}}{\varepsilon_{CP}} \cdot \frac{2(1-v_3) + (1+2v_3)E_3}{(2+v_3) + (1-v_3)E_3} \quad (30)$$

The concatenation of intermediate parameters ends with the inner layer consisting of the nucleus and its envelope.

$$v_3 = \left(1 - \frac{t_{NE}}{r_N}\right)^3 \quad (31)$$

$$E_3 = \frac{\varepsilon_{NP}}{\varepsilon_{NE}} \quad (32)$$

After calculating the effective dielectric constant of the cell according to the above algorithm, the effective material parameters of the mixture are calculated based on the Hanai-Bruggeman equation as presented in [Asami, 2002]

$$0 = \frac{\varepsilon_{mix} - \varepsilon_{cell}}{\varepsilon_{EC} - \varepsilon_{cell}} \sqrt[3]{\frac{\varepsilon_{EC}}{\varepsilon_{mix}}} - (1 - c_{cell}). \quad (33)$$